\UseRawInputEncoding
\documentclass[conference]{IEEEtran}
\IEEEoverridecommandlockouts
\usepackage{cite}
\usepackage{amsmath,amssymb,amsfonts}
\usepackage{algorithmic}
\usepackage{graphicx}
\usepackage{textcomp}
\usepackage{xcolor}
\usepackage{enumitem}
\usepackage{makecell}
\usepackage{subfigure}
\usepackage{stfloats}
\usepackage[linesnumbered,ruled,vlined]{algorithm2e}
\usepackage[T1]{fontenc}
\def\BibTeX{{\rm B\kern-.05em{\sc i\kern-.025em b}\kern-.08em
    T\kern-.1667em\lower.7ex\hbox{E}\kern-.125emX}}
\begin{document}

\title{An Optimal SVC Bitstream Schema for Viewport-dependent 360-degree Video Streaming\\}

\author{\IEEEauthorblockN{Gang Shen}
\IEEEauthorblockA{\textit{NEX} \\
\textit{Intel}\\
Hillsboro, US \\
gang.shen@intel.com}
\and
\IEEEauthorblockN{Mingyang Ma}
\IEEEauthorblockA{\textit{School of Software Engineering} \\
\textit{University of Science and Technology of China}\\
Hefei, China \\
mamingyang@mail.utstc.edu.cn}
\and
\IEEEauthorblockN{Guangxin Xu}
\IEEEauthorblockA{\textit{DCAI} \\
\textit{Intel}\\
Shanghai, China \\
guangxin.xu@intel.com}
}

\maketitle

\begin{abstract}
Abstract¡ª To deliver ultra-high resolution 360-degree video (such as 8K, 12K, or even higher) across the internet, 
viewport-dependent streaming becomes necessary to save bandwidth. 
During viewport switches, clients and servers will instantly exchange coordination info and contents for the given viewports. 
However, those viewport switches pose a serious challenge for video encoding because the temporal dependency 
between contents within changing viewports is unpredictable. In existing practices, it is commonly noted that GOP 
(Group of Pictures) size in a bitstream intrinsically prohibits the reduction of the viewport switch latency, such as 
Motion-to-photon (MTP) latency, or motion-to-high-quality (MTHQ) latency. In this paper, we presented a Scalable Video Coding (SVC) based bitstream schema, which can structurally remove the impacts of GOP in viewport-dependent streaming and provide instant viewport switches within one-frame time (the best possible). In addition, combined with tiling, this new coding schema allows an efficient packing of the non-adjacent regions within a viewport of 360-degree video. Our experiments also show that the overall encoding with this SVC-based approach is faster than with multi-stream approaches. Compared with current 360-degree video streaming solutions based on MPEG-I OMAF, our approach is superior in terms of viewport switch latency, simplicity of viewport packing, and encoding performance.
\end{abstract}

\begin{IEEEkeywords}
360-degree Video, OMAF, Scalable Video Coding (SVC), Motion-to-photon (MTP), Virtual Reality (VR), AV1
\end{IEEEkeywords}

\section{BACKGROUND}
Ultra-high resolution (like 8K, 12K, and even higher) 360-degree video contents are becoming 
desirable in the market. Meanwhile, the bandwidth to support such high-resolution content 
becomes a challenge for the network, even with modern codecs like HEVC or AV1. It is 
estimated that it may take 80-100Mbps to support one 8K 60FPS stream \cite{ichigaya2016required} if encoded in 
HEVC with medium-quality settings. 

While most of the content in a spherical surface is not visible to the client device during consumption, viewport-dependent streaming, in which only content within the client's field of view (FOV) is delivered, intuitively becomes a viable approach (and the only viable approach) for bandwidth-saving and large-scale distributions. In MPEG-I OMAF (Omnidirectional MediA Format), two profiles, advanced video coding (AVC)-based and high-efficiency video coding (HEVC)-based viewport-dependent video profile, as well as three corresponding tile-based streaming approaches, are specified \cite{hannuksela2019overview,choi2017information}. To facilitate viewport-dependent streaming, OMAF specifically uses HEVC MCTS (Motion Constraint Tile Set) in one of the approaches to separate 360-degree video frames into independently decodable tiles, so that contents in any viewports can be packaged, combined, and delivered. Several commercial implementations can be found on different platforms including Android, iOS, Windows, and the Web \cite{o.v.cloudImmersiveVideoSample,tiledmediaHowClearVRDrives,tiledmediaTiledmediaTiledStreaming,podborski2019html5}. Besides, VR video designers will benefit from an OMAF creator system built on the Nokia cloud stream processing platform where end users can upload, edit and preview 6K 360-degree video and watch it with a video link \cite{you2020omaf4cloud}. Although the latest end-to-end system based on OMAF v2 can process and display 8K 60 fps VR video \cite{zhang2022realvr}, there are challenges like heavy computations, complex file packaging, and viewport switch latency. 
This paper introduces an optimal bitstream schema for 360-degree video, based on SVC, to cope with current challenges in OMAF. Particularly, it offers an unparallel answer to the interactive latency during the viewport switch, which is critical and common for viewport-dependent streaming. In the following, section II illustrates challenges in 360-degree video format and streaming; section III gives the design of this SVC-based new coding schema; section IV provides a reference implementation by AV1 SVC; section V shows encoding experiments and performance analysis; section VI gives conclusions.

\section{THE CHALLENGES IN VIEWPORT-DEPENDENT 360-DEGREE VIDEO STREAMING}

\subsection{Viewport Change and MTP/MTHQ Latency}
\begin{figure}[htbp]
    \centerline{\includegraphics[width=0.5\textwidth]{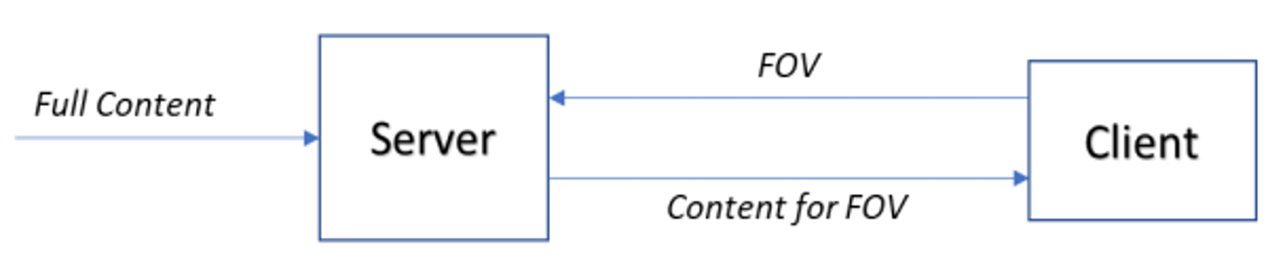}}
    \caption{The exchange of viewport and content in viewport-dependent streaming}
    \label{exchange}
\end{figure}

The success of viewport-dependent streaming relies on the efficiency of the exchange between 
viewport information and partial content to cover that viewport. While this exchange (shown in Fig. \ref{exchange})
 saves significant bandwidth, it introduces an interactive latency between server and client, 
 which can tamper the visual quality during viewport changes. In addition, to support large-scale 
 distribution, the server in Fig. \ref{exchange} cannot feasibly rely on real-time transcoding to prepare streams 
 for individual and dynamic viewports of many clients. It is necessary to have an optimal content 
 (bitstream) schema which can support ultra-low latency for exchange and scalability for distribution.

The exchange/interactive latency is defined in MPEG-I as MTP (Motion-to-Photon) 
latency when the single quality of content is used, or MTHQ (Motion-to-High-Quality) 
latency in the case of two levels of resolutions/qualities of contents are used. 
Human eyes are sensitive to detect visual quality gaps - according to \cite{albert2017latency}, this latency 
is required to be as little as 50ms to give users completely immersive experiences, 
which VR or 360-degree video products in today's market can hardly achieve.

There are known methods to reduce this latency in the industry, based on either video encoding, 
packing, or network transport. The popular approach as for now, in MPEG-I OMAF, 
provides a combined approach of multi-bitrate or multi-resolution packing, 
region-wise packing (RWP), bitstream rewriting, and tile-based coding to offer a bandwidth-saving solution \cite{hannuksela2019overview}. However, it can only mitigate this interactive latency by adding additional encoding tracks of long and short Group of Pictures (GOP) sizes - i.e., during the viewport change, the content will be selected and transported from short GOP tracks. But this approach will increase the transcoding workloads and bandwidth for transporting 
short GOP track.

\begin{figure}[htbp]
    \centerline{\includegraphics[width=0.5\textwidth]{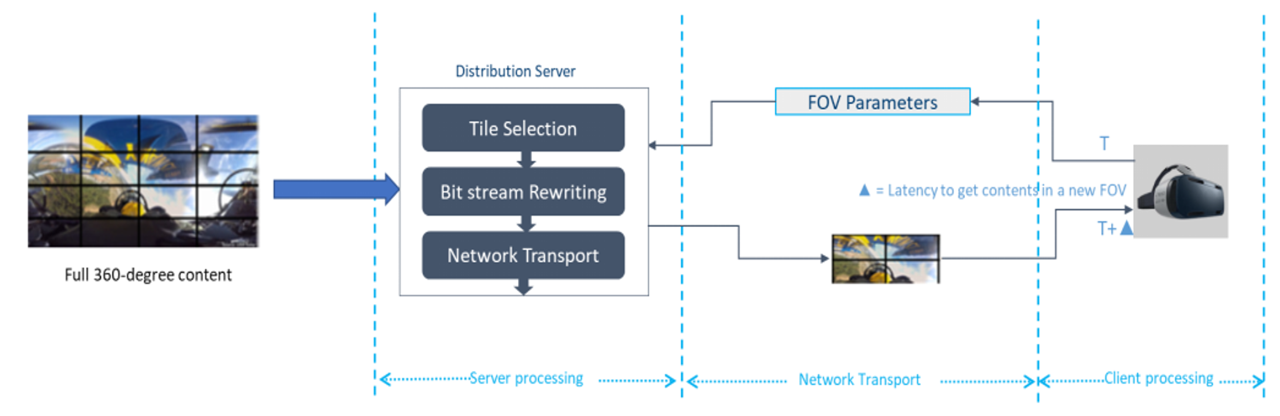}}
    \caption{An example implementation of a distribution server for viewport-dependent 360-degree video streaming}
    \label{distrubute}
\end{figure}

When a viewport change happens in live VR/360 video viewport-dependent streaming, 
the access to the content changes both spatially - from one spot to another spot, 
and temporally - from this frame to the next frame. The distribution server (Fig. \ref{distrubute}) - usually at the edge network, which will receive full content from the source, and need to select partial contents, according to the FOV (sent by the client), within the bitstream(s). 
However, the selection of partial contents can occur meaningfully with IDR frames - the first frame of GOP. The longer GOP structure, the longer delay to make the viewport change; Meanwhile, the shorter GOP, the higher bitrate. Therefore, the GOP size is a difficult factor to decide, in commercial implementations of viewport-dependent 360 video live streaming.

\subsection{Projections and Packing of Viewport Contents}
Meanwhile, the 360-degree video needs specific projections and packing methods to be 
encoded in 2D rectangular frames for encoding and network transport. Due to the projections 
from a spherical space to a planar space, the contents for a given FOV are not always 
adjacent in 2D frames. This will cause difficulties and complexity for collecting, arranging, 
and packing contents of FOV into one rectangular, which is the usual requirement of 
the interfaces of either encoders or decoders.

\begin{figure}
    \centerline{\includegraphics[width=0.5\textwidth]{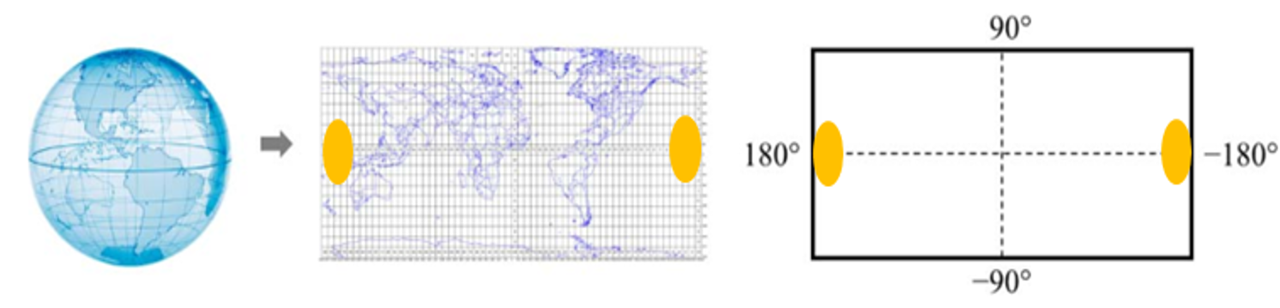}}
    \caption{Equirectangular projection and packing in MPEG-I OMAF (\cite{choi2017information})}
    \label{projection}
\end{figure}

For example, in equirectangular projection (shown in Fig. \ref{projection}), a field of view could be composed of nonadjacent regions (marked by orange ovals) by the left and right edges.

\begin{figure}
    \centerline{\includegraphics[width=0.5\textwidth]{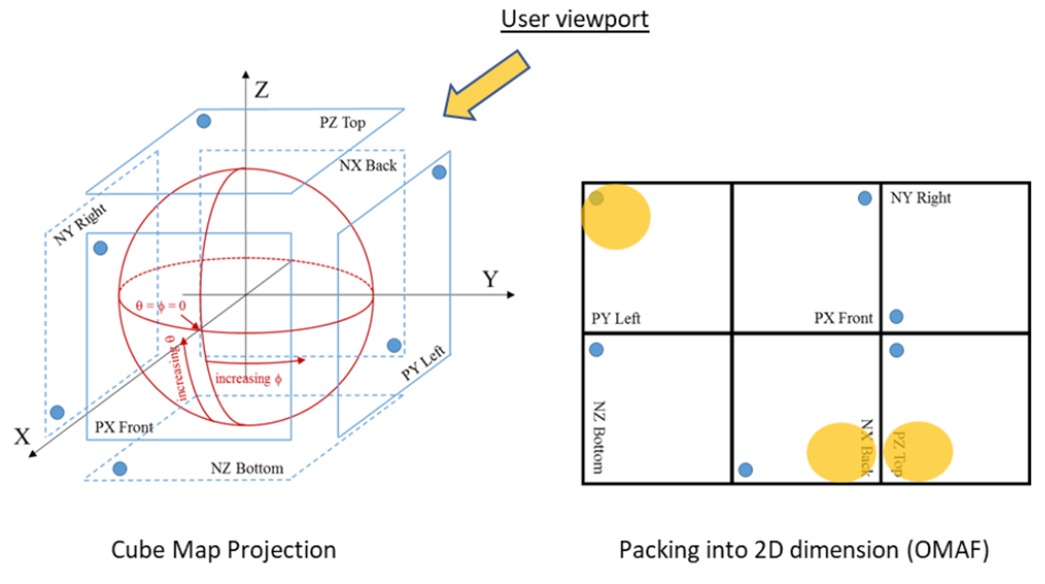}}
    \caption{Cube map projection and packing in MPEG-I OMAF (\cite{choi2017information})}
    \label{cubemap}
\end{figure}

Cube map projection has the same issue. As shown in Fig. \ref{cubemap}, suppose (1) the given 
viewport is coming top-right-back corner; (2) the content sphere is projected to a cube; 
(3) the six surfaces of the cube are packed into a 2D rectangular based on 
OMAF specification \cite{choi2017information}, the contents for this viewport are distributed in nonadjacent regions marked by orange ovals.

Those nonadjacent regions of one given viewport are challenges for region-wise packing (RWP), tile-based encoding on the server side, as well as late-binding for decoders in client devices. 
Even with tile-based video encoding (like HEVC), it still needs a delicate bitstream rewriting algorithm, specified in MPEG-I OMAF v2.0(\cite{choi2017information}, 4.2.3), to pack those nonadjacent regions into a 2D rectangular or a conformant bitstream(s) for single or multiple decoders on the client devices. 
Packing multiple regions into one rectangular is geometrically difficult, and costly 
when using paddings. 

\section{AN OPTIMAL SVC-BASED BITSTREAM SCHEMA}
Based on discussions in section II, the viewport-dependent 360-degree video streaming 
faces challenges from the temporal domain of video coding - GOP size's 
impact to interactive latency, as well as from the spatial domain of video coding - 
nonadjacent regions of a given FOV. It suggests that an advanced coding schema may 
be needed for 360-degree video.

In fact, conventional video coding also faces challenges to adapt to various demands 
of the user ends and enhance encoding performance, where the SVC and tiles are developed respectively. 
The main concept of SVC is to provide partially removable video streaming by separating it into 
a base layer and enhanced layers. Tiles divide a video frame into independent parts and a tile is 
a rectangle of superblock whose spatial referencing is limited to be within the tile boundary. 
Different tiles within a frame can be encoded separately which facilitates multi-threading capabilities.

\begin{figure}
    \centerline{\includegraphics[width=0.5\textwidth]{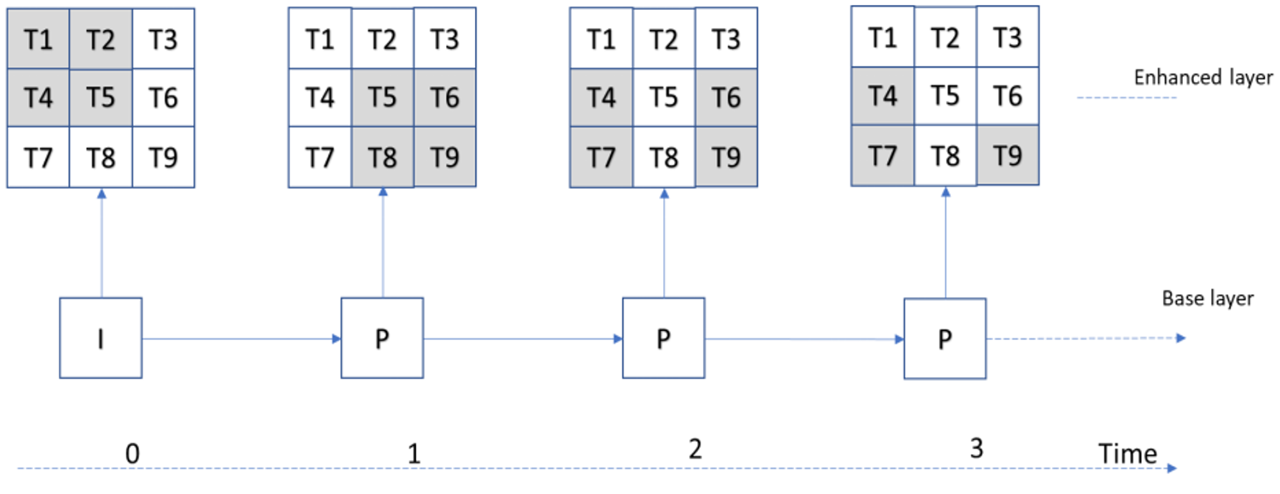}}
    \caption{An optimal coding schema for viewport-dependent 360-degree video streaming}
    \label{pic5}
\end{figure}

Therefore, we come up with an optimal video coding schema (Fig. \ref{pic5}) based on SVC and Tiles, 
which will overcome the difficulties of conventional coding used in MPEG-I OMAF.

There are two key elements in this new coding schema: (1) to use enhance layer in SVC for 
high-quality contents while using base layer for low-quality contents; (2) to use tiles 
(no coding dependency between tiles) segment one frame into multiple regions.

\begin{figure}
    \centering
    \subfigure[Enhanced layer refers to one frame]{
        \label{fig6_1}
        \includegraphics[width=0.45\textwidth]{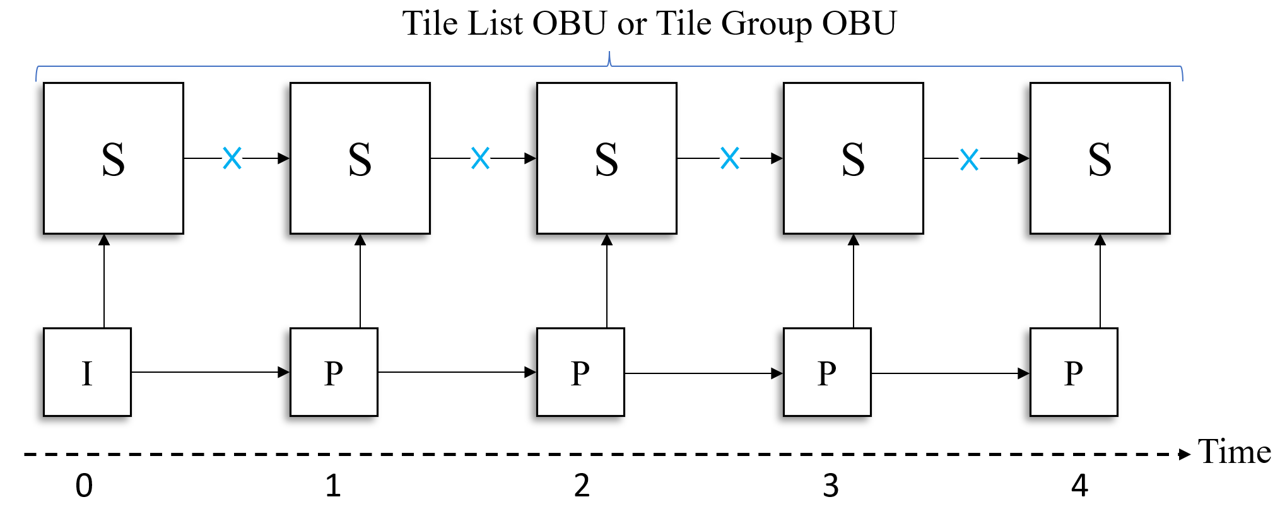}}
    \subfigure[Enhanced layer refers to mutilple frames]{
        \label{fig6_2}
        \includegraphics[width=0.45\textwidth]{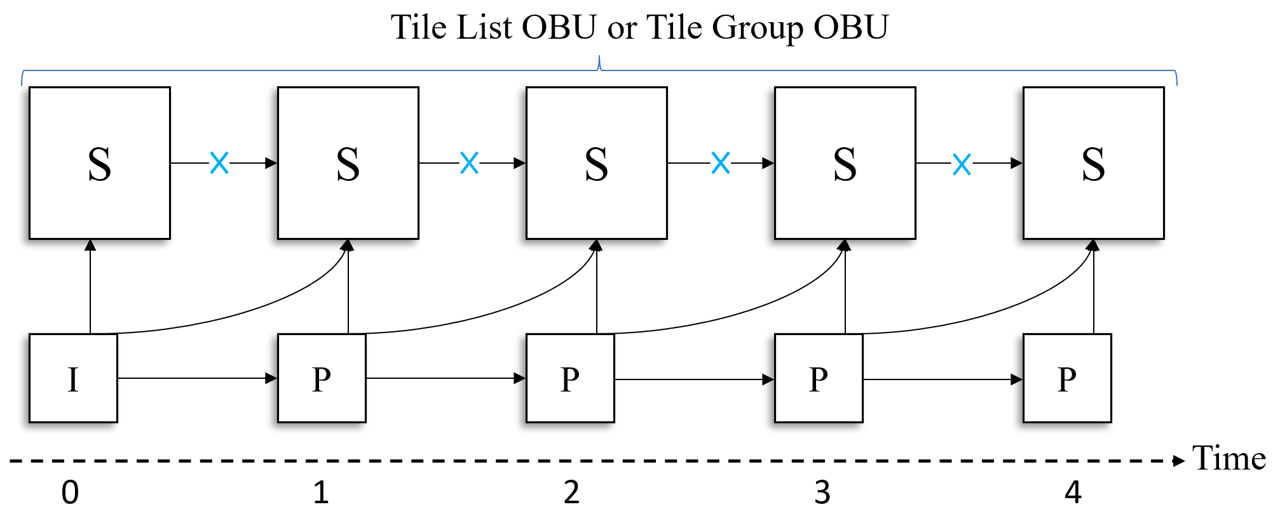}
    }
    \caption{L2T1 structure in AV1 SVC and Enhancements}
    \label{fig6}
\end{figure}

In MPEG-I OMAF, AVC-based and HEVC-based viewport-dependent OMAF video profiles combine 
high-quality contents of FOV with full view of low-quality contents using multiple streams (Fig. \ref{fig6}). 
The low-quality contents will be the backup when high-quality contents cannot cover the FOV when 
rendering on client devices. Comparably, the base layer of SVC can be the backup and the tiles of 
enhanced layer can compose the contents (even nonadjacent regions) for any given FOVs. 

Furthermore, the enhanced layer can have only spatial dependency on base layer. 
This implies the only temporal dependency in this structure is on base layer. 
Since all frames in base layer will be transported to client, the high-quality tiles in 
enhanced layer can be accessed freely without the constraint of GOP.

The detailed design of the coding schema is listed as following:
\begin{enumerate}
    \item It has two layers: low-resolution (or low-quality) as base layer and high-resolution 
    (or high-quality) as enhanced layer. It is possible to extend to multiple enhanced layers 
    (for example, for Simulcast). Also, the base layer can have temporal dependency - i.e., 
    it can have I-frame, P-frame and GOP (Group of Pictures), so that the video compression 
    benefits can be achieved. The GOP in this proposal refers to \lq\lq Close GOP\rq\rq.
    \item Frames in enhanced layer will have spatial dependency on based layer only. 
    In the Fig. \ref{pic5}, the dependency between enhanced layer and base layer is on the same frame, 
    but it is viable for frames in enhanced layer have this dependency on a few previous frames 
    in base layer: for example, the frame 3 in enhanced layer may have dependencies in 
    frame 1 or 2 in base layer. The key is, keep temporal dependency between frames out of 
    enhanced layer.
    \item Frames in the base layer will have a temporal dependency on the multiple previous 
    based layer frames. For example, frame 3 in the base layer may have dependencies on frame 
    1 or 2 in the base layer.
    \item The base layer may only have a single tile/slice to save bitrate.
    \item When streaming to a client device, only content (grey tiles) within the given FOV in 
    enhanced layer and full tiles in base layer will be delivered. The location of each tile 
    will be carried in manifest message in bitstream (like SEI message \cite{sullivan2012overview} in HEVC or OBU 
    header in AV1) of each frame.
\end{enumerate}

During viewport changes on 360-degree video (or panorama video) streaming, this coding structure 
can offer immediate and random access to any tiles in the enhanced layer. The nonadjacent regions 
of a viewport can be overcome if bitstream can accommodate the position information of each tile, 
then no need for an additional packing method like that defined in MPEG-I OMAF.

The grey tiles in Fig. \ref{pic5} indicate contents/tiles for a given FOV, respectively at time 
0, 1, 2, 3. Certainly, FOV may not change so drastically from frame to frame. This is for 
illustration of the viability and flexibility of this bitstream schema:

\begin{enumerate}
    \item From frame 0 to frame 1, new FOV needs to decode new T6, T8, T9 tiles 
    (T5 exists in Frame 0; T5 is decodable). Since those tiles depend on base layer only, 
    those can be decoded.
    \item Frame 2 shows that the content for a new FOV includes nonadjacent tiles: 
    T4, T6, T7, T9. It is still feasible if SEI message (or header information) can provide 
    the position info of each tile.
    \item Frame 3 shows that it allows to have different quantity of tiles in this structure. 
    It is useful for some formats (e.g., ERP) of 360-degree video where t the projection may 
    make the uneven distribution of pixels for viewports.
\end{enumerate}

So, by weaving the contents into tiles and layers, this coding schema removes both 
the constraints of GOP structure and the need of region packing. Furthermore, 
tiles and layers are built-in concepts or tools in modern codecs like AV1 and VVC. 
Based on this coding schema, the end-to-end implementations of 360-degree video streaming 
solutions can be greatly simplified.

\section{REFERENCE IMPLEMENTATION USING AV1 SVC}

This structure is implementable by modern codec standards. For example, in AV1 SVC, 
the structure can be built upon L2T1 \cite{de2018av1}, with additional tile specifications.

Fig. \ref{fig6} is L2T1 scalability structure in AV1 SVC. To support the structure in this proposal, 
there are a few extensions to make:
\begin{enumerate}
    \item Remove the temporal dependency between frames in enhanced layer, marked by ¡°X¡± in the Fig. \ref{fig6}.
    \item The base layer can be encoded as a single tile to save bitrate. 
    The enhanced layer needs to be encoded as multiple tiles to enable a quick viewpoint switch.
    \item Use \lq\lq Tile Group OBU\rq\rq or \lq\lq Tile List OBU\rq\rq to select and pack tiles for a given FOV. 
    Based on the current AV1 SVC Spec, \lq\lq Tile Group OBU\rq\rq is more reasonable for this case.
    \item Each tile should be packed or delimited independently - it will help 
    independent handling of high-resolution tiles in the transport layer.
    \item Enhanced layer can refer to one frame in the base layer like Fig. \ref{fig6_1} or 
    multiple previous frames in the base layer like Fig. \ref{fig6_2}.
\end{enumerate}

\begin{figure}
    \centerline{\includegraphics[width=0.5\textwidth]{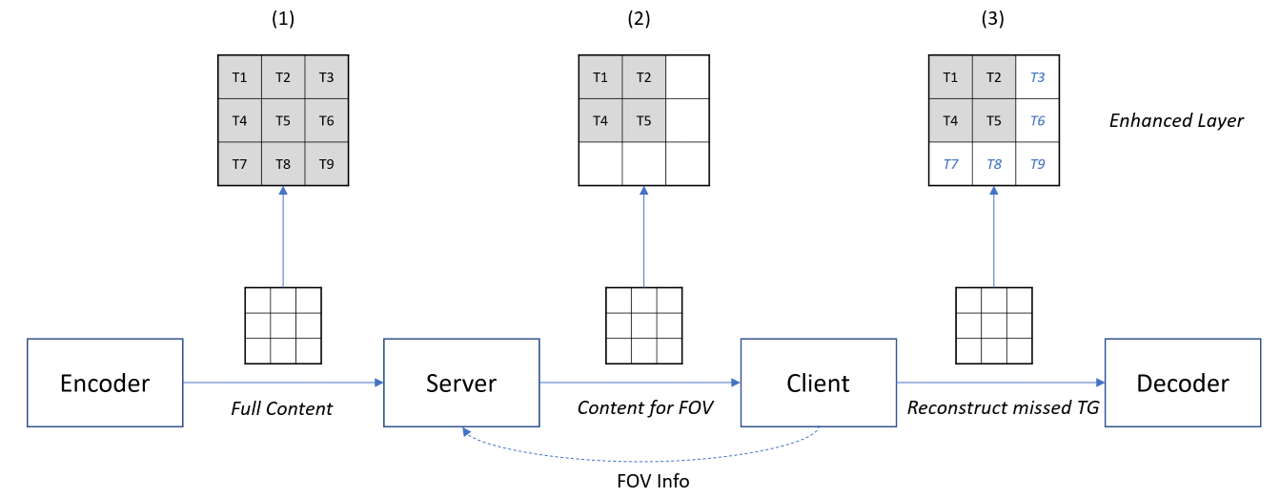}}
    \caption{The proposed end-to-end bitstream processing}
    \label{end2end}
\end{figure}

\begin{figure}
    \centerline{\includegraphics[width=0.5\textwidth]{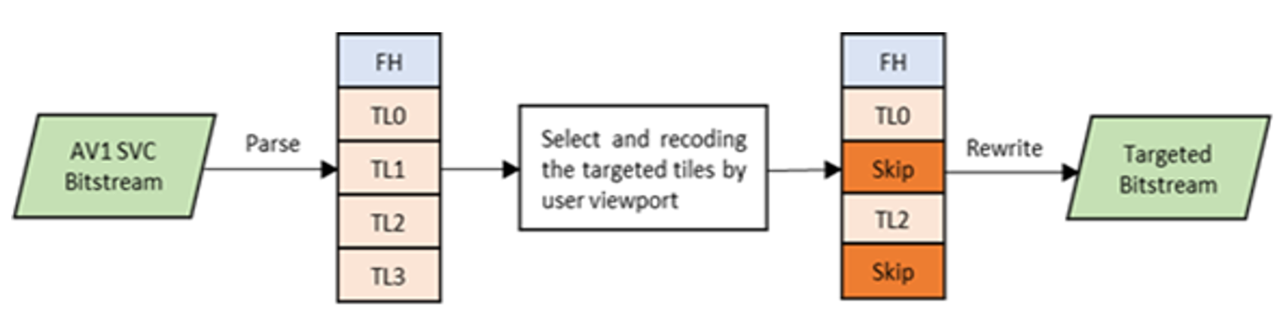}}
    \caption{The overview of rewriter on AV1 bitstream OBU structure}
    \label{av1overview}
\end{figure}

Here, a reference implementation to show how to use \lq\lq Tiled Group OBU\rq\rq to support the 
viewport switch and deliver content for FOV is shown in Fig. \ref{end2end}:
\begin{enumerate}
    \item The encoder will encode a high-quality layer for the entire picture, 
    with tiling to split full content into small regions. According to AV1 SVC, 
    multiple Tile Group OBUs can be applied on those tiles and in this case, 
    each Tile Group OBU will cover one and only one tile - the \lq\lq T\#\rq\rq in Fig. \ref{end2end} refers to one tile. 
    The base layer has the same tiling on lower resolution or lower quality - 
    it does not affect the algorithm if the base layer has a single tile. 
    The output of the encoder is illustrated as (1) in Fig. \ref{end2end}.
    \item The server will receive full content as the output of encoder (1); 
    Meanwhile, the server will also receive viewport information (FOV info) from clients - 
    only one client is shown in Fig. \ref{end2end}.
    \item The server will apply the tile selection and repackaging (bitstream rewriting) process 
    on the full content frames (1), according to the viewport info from a given client. 
    The tile selection will be a mapping from the viewport parameter (Fig. \ref{av1overview}) to a list of tiles 
    within the viewer\rq s FOV - For example, T1, T2, T4, and T5 in (2) in Fig. \ref{end2end}. 
    Then, the server will extract those tiles (based on the syntax of AV1 SVC) and repackage 
    them into a new \lq\lq frame\rq\rq with only tiles within FOV in the enhanced layer and all tiles 
    in the base layer.
    \begin{enumerate}
        \item Since every tile is inside one Tile Group OBU and the Tile Group OBU has no other tiles, 
        the tile\rq s position is identified by\lq\lq tg\_start\rq\rq and \lq\lq tg\_end\rq\rq values in OBU.
        \item After the repackaging of selected tiles, a temporal delimiter OBU should be 
        applied which will identify the boundary between frames.
        \item The output of this step - the content for FOV, is illustrated by (2) in Fig. \ref{end2end}. 
        The tiles outside FOV are skipped.
    \end{enumerate}
    \item Once the client receives all high-resolution tile groups (greyed tiles), 
    it can generate unselected and unreceived tiles as \lq\lq skipped\rq\rq tiles 
    (tiles in blue color, in (3) of Fig. \ref{end2end}), from the base layer, as long as:
    \begin{enumerate}
        \item At frame-level, CDF (Cumulative distribution function) update is set to 
        being disabled and global MV (Motion Vector) is set to zero.
        \item At the tile level, the transform skip param is set to true, 
        and \lq\lq use global MV\rq\rq is set to true.
    \end{enumerate}
    \item According to AV1 SVC specification, those generated \lq\lq skipped\rq\rq tiles are decodable. 
    After decoding, the regions of skipped tiles in the enhanced layer will be filled by taking and scaling up the counterparts from the base layer. There is no change needed on the decoder.
\end{enumerate}

Then, a bitstream rewriter can be implemented to select and construct a \lq\lq viewport-dependent frame\rq\rq 
(shown as (2) in Fig. \ref{end2end}) accordingly. It will rewrite each target tile in each enhanced layer. 
Each superblock in the target tile will be rewritten to the same structure. 
According to the AV1 specification, the syntax of superblock should be rewritten as followings:
\begin{enumerate}
    \item The \textbf{partition} syntax is written as None.
    \item The \textbf{skip} syntax is written as True for no transform coefficients.
    \item \textbf{is\_inter} syntax is True and it only refers to the base layer.
    \item The motion vector is set as zero.
    \item As required by the specification, \textbf{use\_obmc} syntax is set to False to use the simple translations.
\end{enumerate}

Here are pseudo codes for the bitstream rewriter, to create a \lq\lq viewport-dependent frame\rq\rq (2) in Fig. \ref{end2end}: 

\begin{algorithm}
    \caption{Bitstream Rewriter}
    \ForEach{ enhance layer}{
        \ForEach{ target tile}{
            \ForEach{ superblock}{
                \textit{rewrite\_sb(\textbf{superblock})}
            }}}
\end{algorithm}

\begin{algorithm}
    \caption{Rewrite\_sb}
    \textit{write\_partion\_mode(\textbf{PARTITION\_NONE})}\\
    \textit{write\_skip(\textbf{True})}\\
    \textit{write\_is\_inter(\textbf{True})}\\
    \textit{write\_ref\_frames(\textbf{REF\_TO\_BASE\_LAYER\_ONLY})}\\
    \textit{write\_inter\_mode(\textbf{zero\_mv})}\\
    \textit{write\_use\_obmc(\textbf{False})}
\end{algorithm}

\begin{figure}
    \centerline{\includegraphics[width=0.5\textwidth]{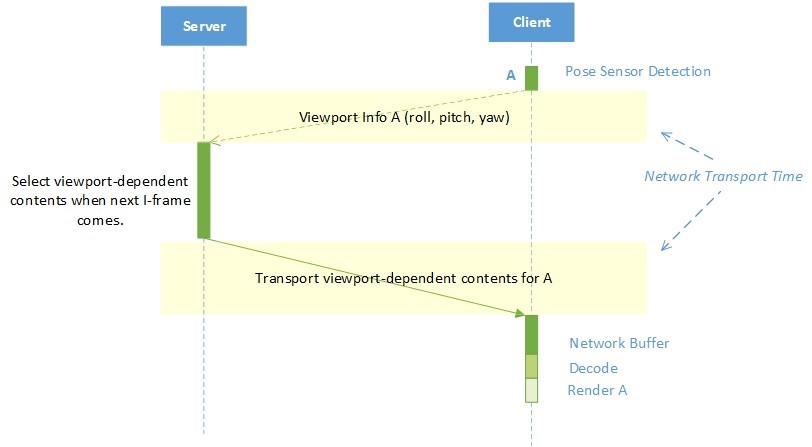}}
    \caption{The overview of rewriter on AV1 bitstream OBU structure}
    \label{switch}
\end{figure}

Fig. \ref{switch} shows that how GOP size affects MTP/MTHQ latency. The entire viewport switch process includes 
several different factors: such as pose detection, network transport, buffering, decoding and rendering. 
Among those, GOP size is a structural factor, which forbid immediate viewport-dependent content selection 
on non-I-frames. Without layered bitstream structure like that in the paper, server will have to wait for 
next I-frame to respond viewport change and it will cost the time about half of GOP size on average. 
For example, if GOP size is 10 and FPS (frames per second) is 30, latency caused by GOP size will be 167ms. 
Although this latency can be reduced by a separate track with short GOP size (e.g., using GOP size 3 or 5), 
such track will increase drastically the bitrate and encoding complexity.

\section{EXPERIMENTS}
\subsection{Environment Setup}
We extended the AV1 codec of Alliance for Open Media (AOM) to implement this schema and made comparisons 
with existing approaches for 360-degree videos. While the new coding schema can achieve instant viewport 
change and much more advanced than existing approaches (as shown in section 3.1), we would like to further 
evaluate this new coding schema in two aspects: (1) quality with the same bitrate; (2) encoding time/speed.

An Intel Xeon E5-8280 CPU and 192GB memories are used to handle 8K and 4K video encoding. 
Three 30fps 8K and 4K 360-degree video clips (in Table I) are used in experiments; 
ERP screenshots are presented in Fig. \ref{clips}. The encoding parameters are set as the following: 
\textit{cpu-used=7, passes=1, threads=20, end-usage=CBR} and \textit{real-time} mode. Each case will be tested three times 
and the average of all the experimental results are used.

\begin{table}[htbp]
    \caption{Video Clips Information}
    \begin{tabular}{|c|p{10em}<{\centering}|c|c|c|}
    \hline
    \textbf{Name}&\textbf{Description}&\textbf{Length}&\textbf{Source}&\textbf{Resolution}\\
    \hline
    Tokyo&People in Tokyo Street&23s&\cite{insta360Insta360Pro8K}&7680x3840\\
    \hline
    Office&People in an Office&41s&\cite{insta360Insta360Pro8K}&7680x3840\\
    \hline
    Beach&Overlooking Caribbean Beach&60s&\cite{airpianoCaribbeanParadise360}&7680x3840\\
    \hline
    Lion&Lions in the Wild&30s&\cite{geographicLions360}&3840x1920\\
    \hline
    Ski&Ski Record&30s&\cite{airpianoRosaKhutorSki}&3840x1920\\
    \hline
    Street&Street Scene&34s&\cite{insta360Insta360Pro8K}&3840x2400\\
    \hline
    \end{tabular}
\end{table}

\begin{figure*}[htbp]
    \centering
    \subfigure[Toyko]{
        \includegraphics[width=0.32\textwidth]{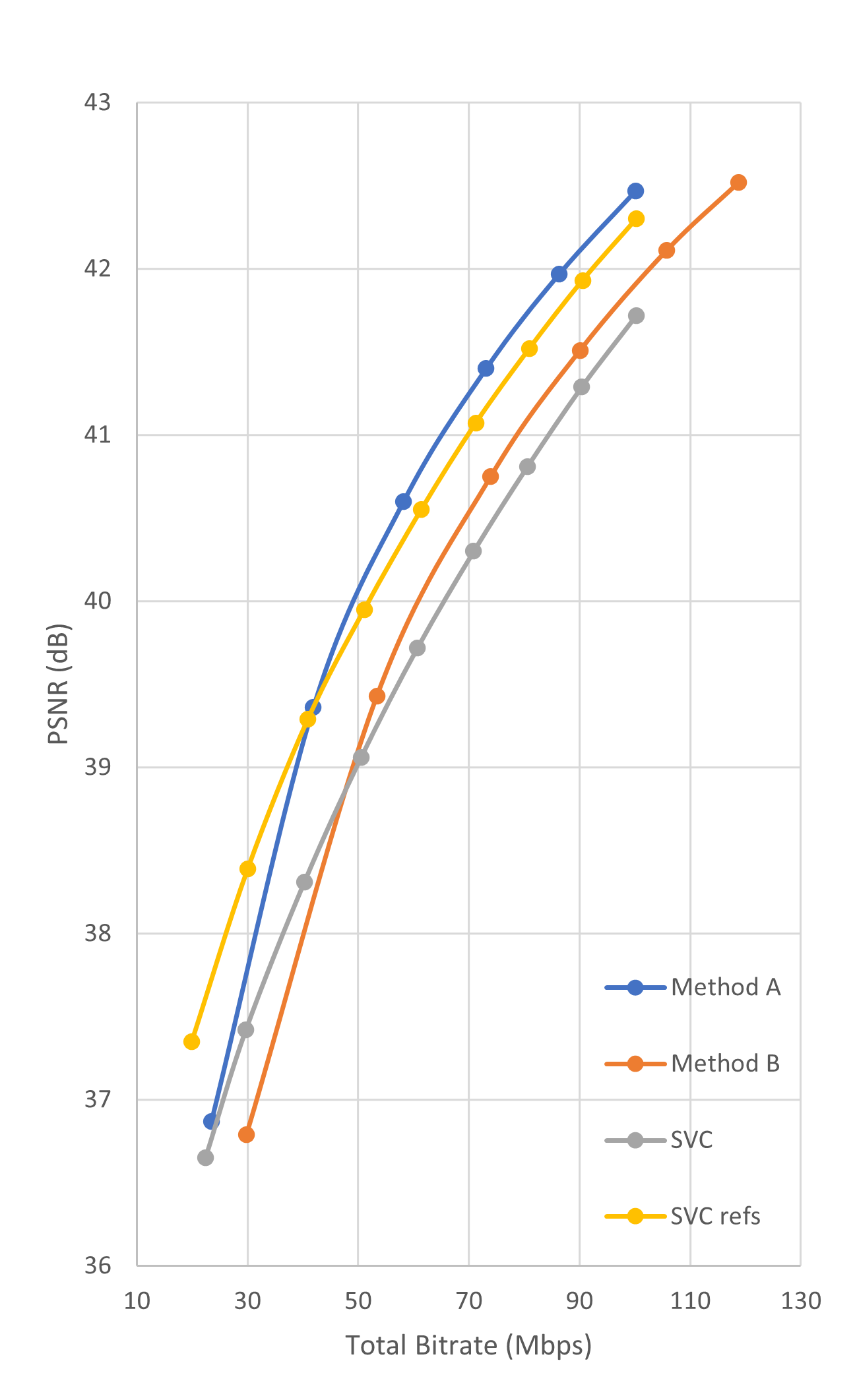}}
    \subfigure[Office]{
        \includegraphics[width=0.32\textwidth]{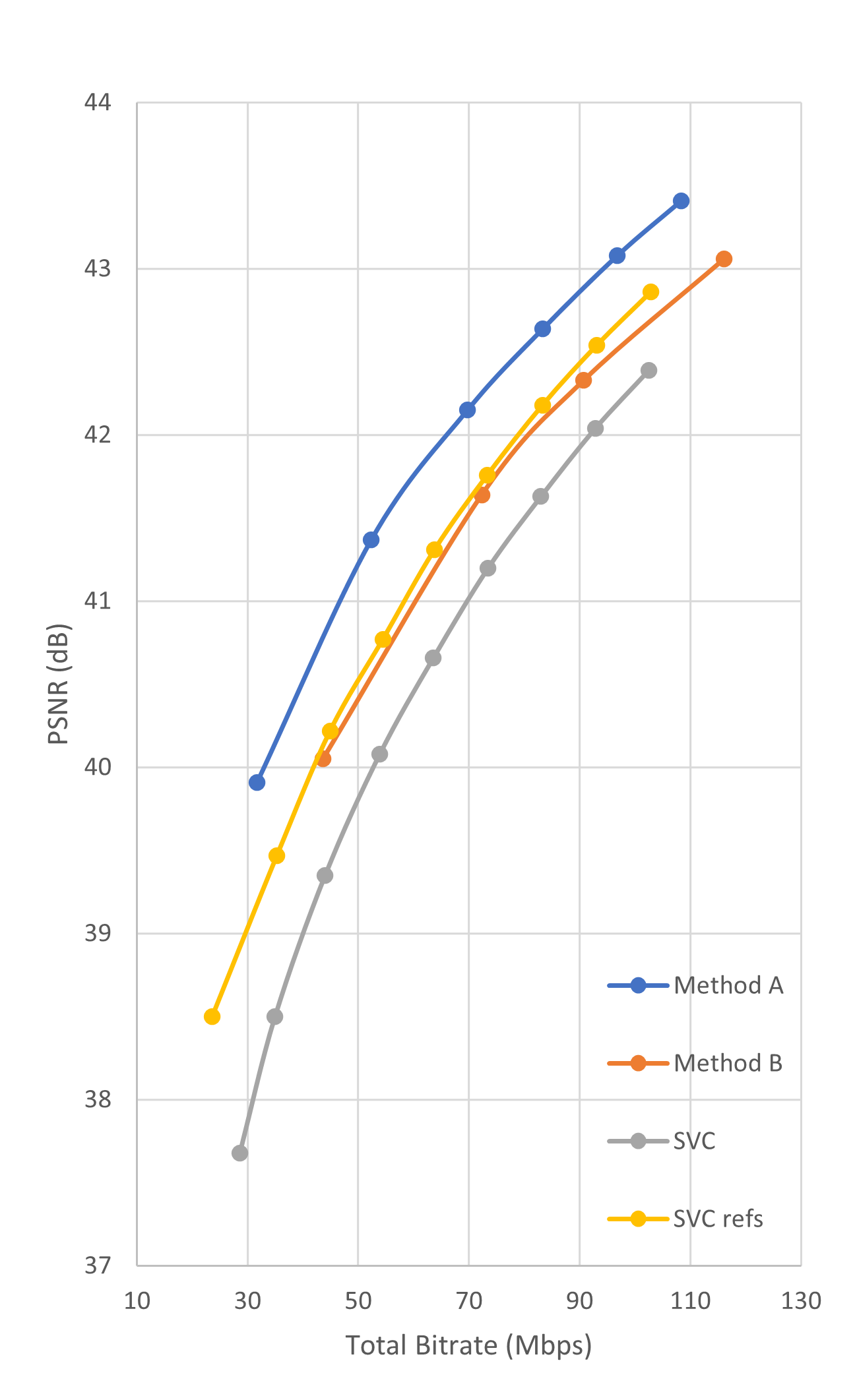}}
    \subfigure[Beach]{
        \includegraphics[width=0.32\textwidth]{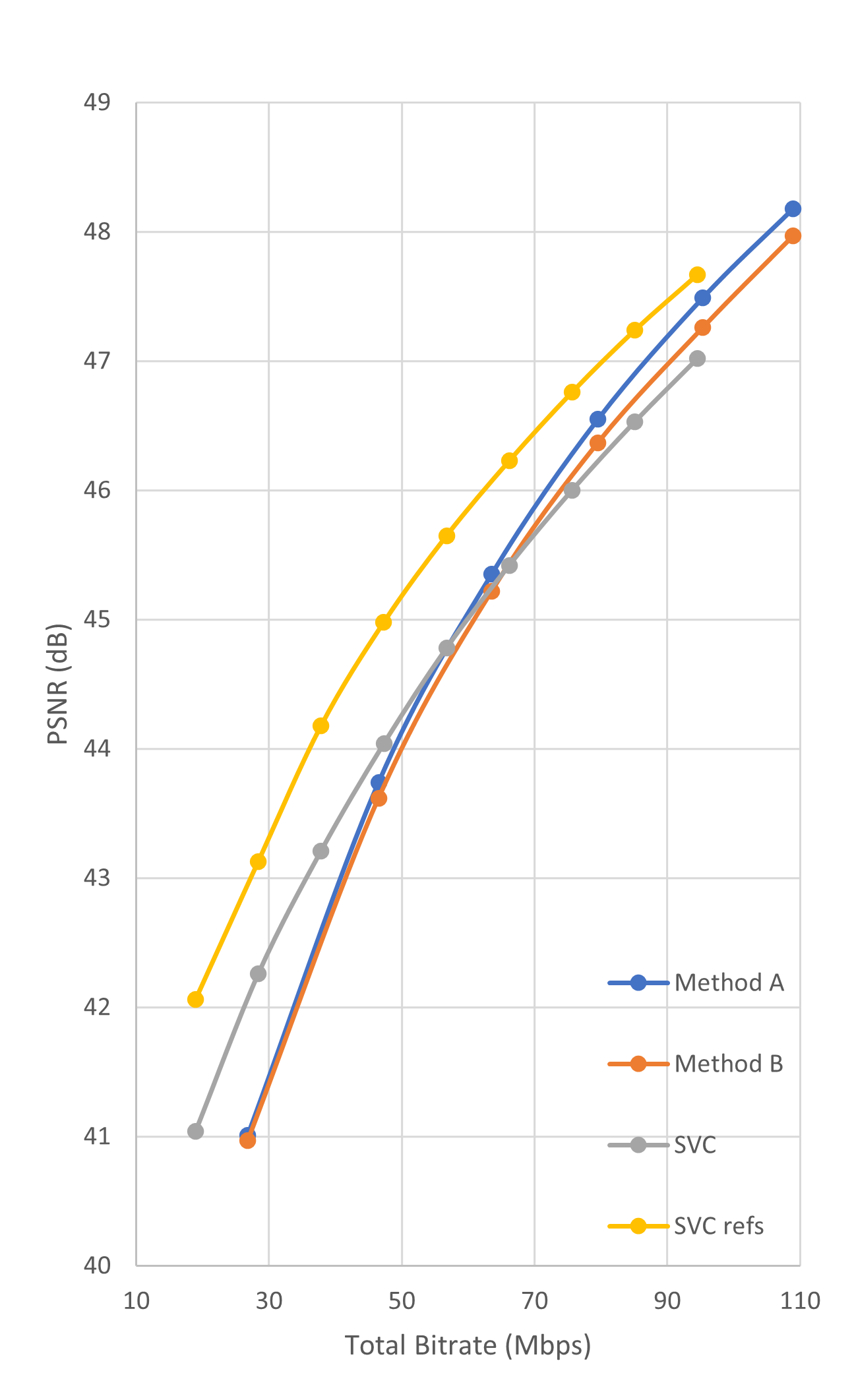}}
    \subfigure[Lion]{
        \includegraphics[width=0.32\textwidth]{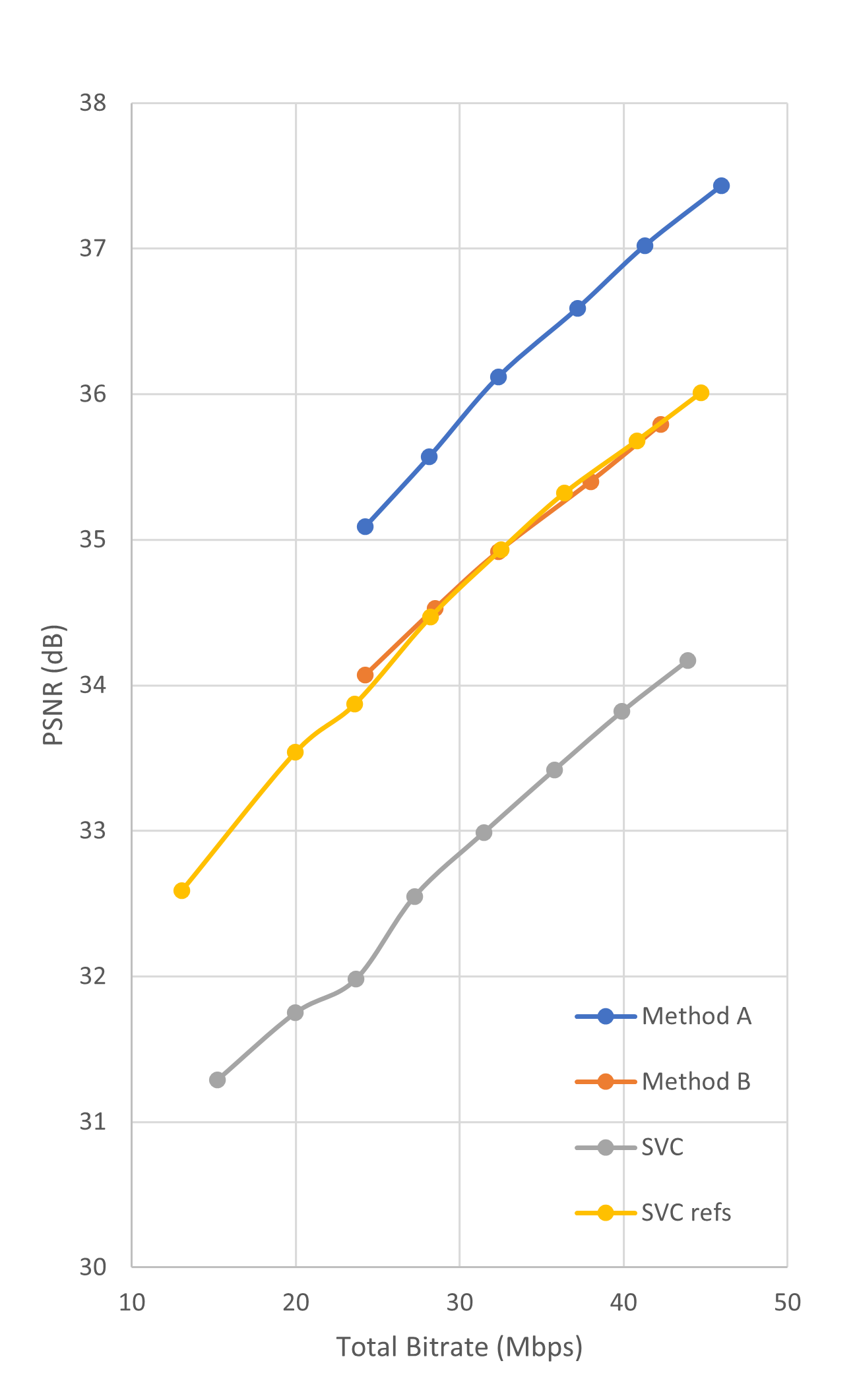}}
    \subfigure[Ski]{
        \includegraphics[width=0.32\textwidth]{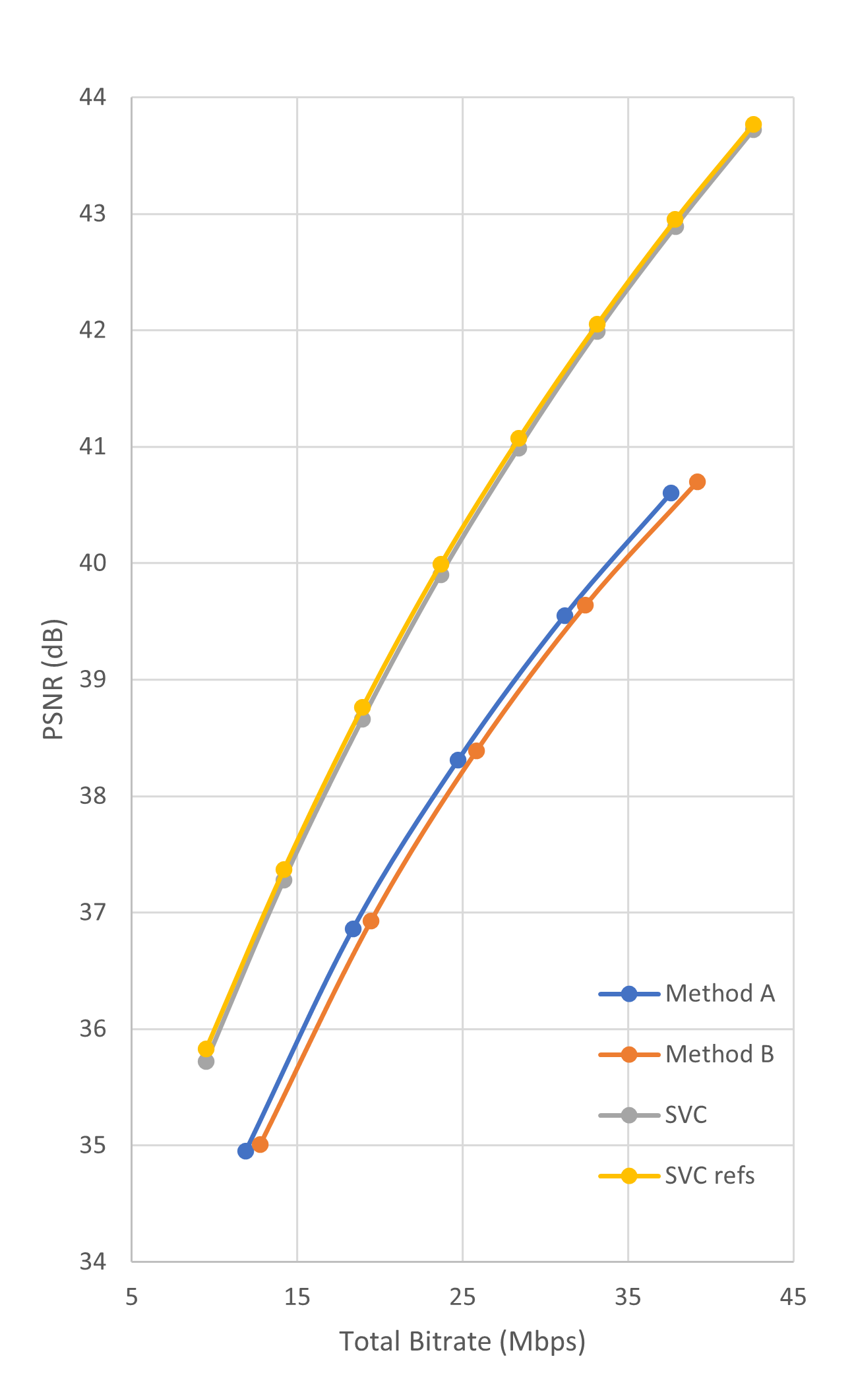}}
    \subfigure[Street]{
        \includegraphics[width=0.32\textwidth]{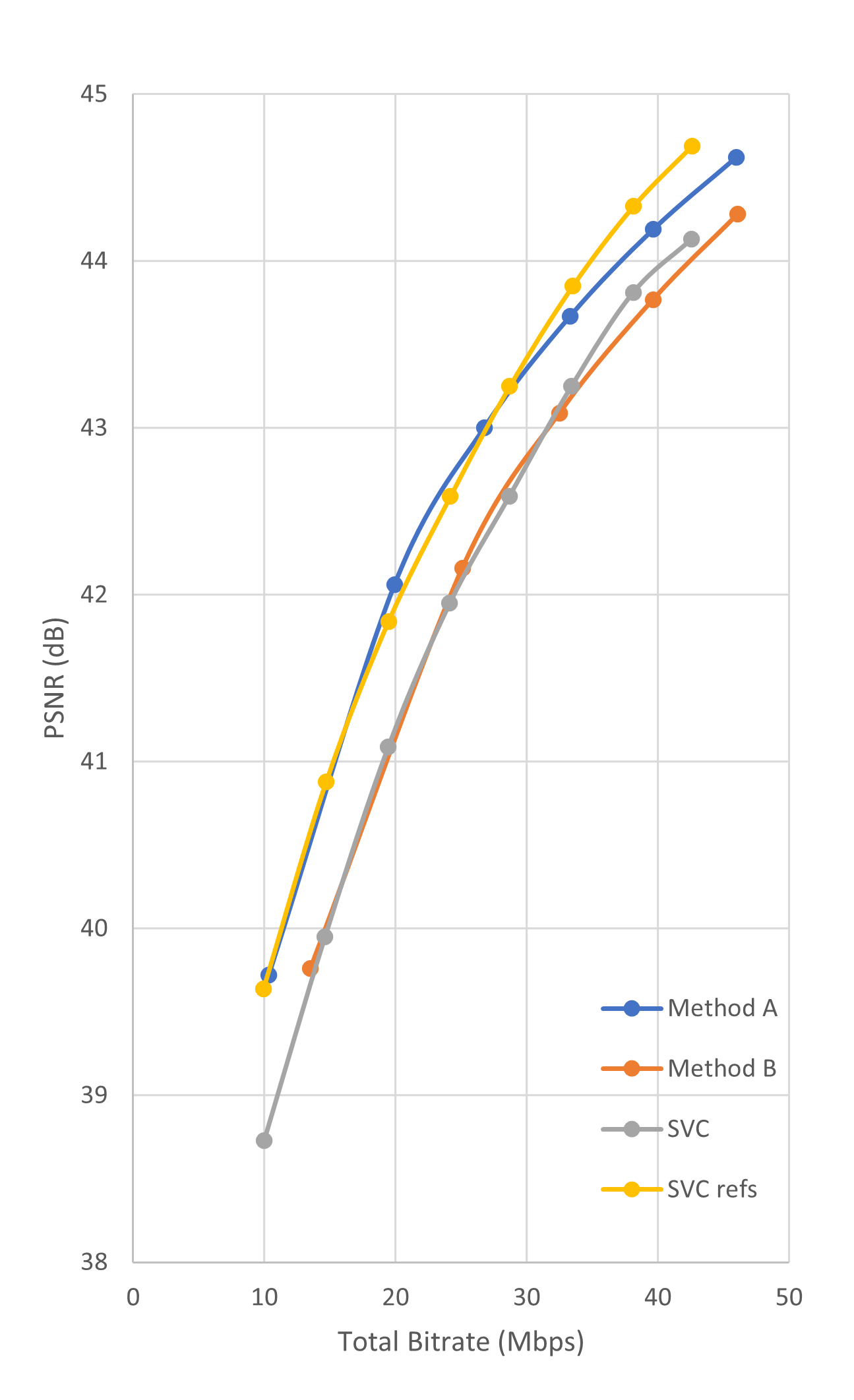}}
    \caption{PSNR under different encoding configurations}
    \label{PSNR}
\end{figure*}

\subsection{Bitrate and Quality}
In industry, the 360-degree video solution usually includes two or three streams, e.g., one high resolution track with long GOP size for long time view, one low resolution track with 
long GOP size as background and one high resolution with short GOP size for viewport switch. 
To make the evaluation meaningful, we compare PSNR of two methods in selected video clips. 
For existing solution, the PSNR is the average of two high resolution tracks and the scalable ratio is 2.0x, which means the low quality video is half the width and height of the original video.

\begin{figure}
    \centerline{\includegraphics[width=0.5\textwidth]{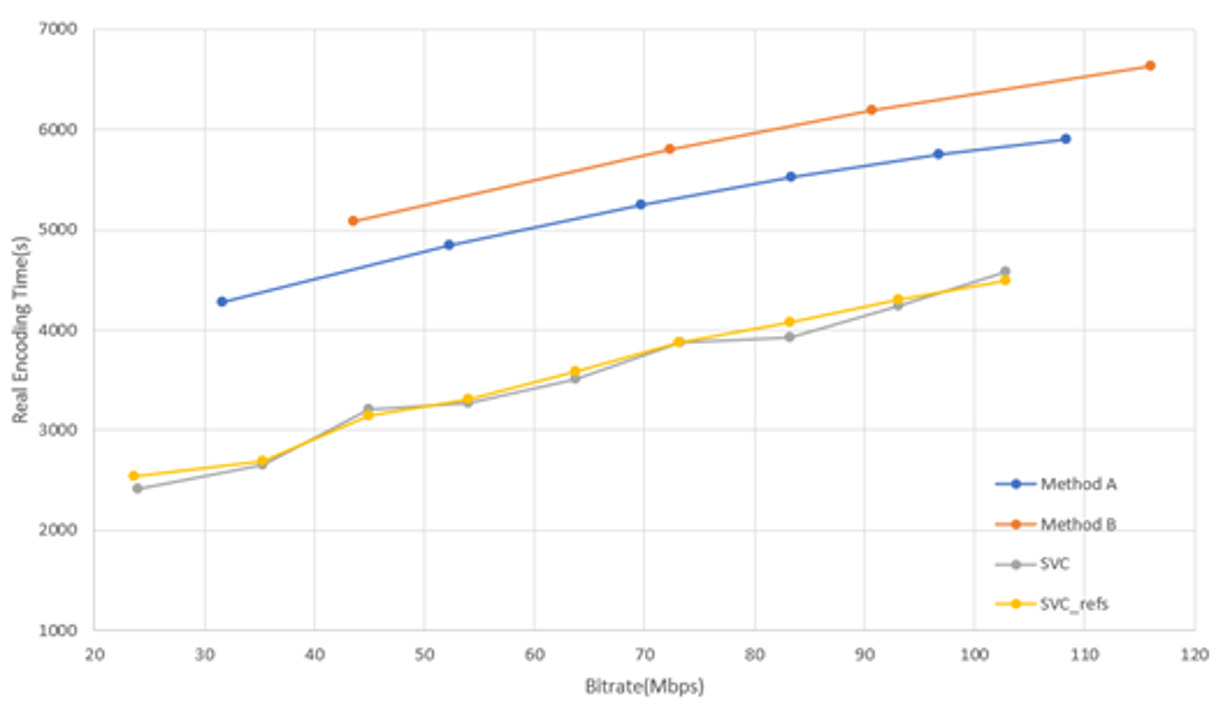}}
    \caption{Encoding time under different configurations}
    \label{enc_time}
\end{figure}

In Fig. \ref{PSNR} \& \ref{enc_time}, method A and B are common three-track methods stated as above, 
and the long GOP size is 30 and the short GOP size is 5 and 3, respectively. 
The new SVC-based coding schema includes two structures in Fig. \ref{fig6}: enhancement layer refers 
only one frame and multiple frames of base layer (denoted as \lq\lq SVC\rq\rq and \lq\lq SVC refs\rq\rq respectively 
in Fig. \ref{PSNR} \& \ref{enc_time}).

Fig. \ref{PSNR} shows that SVC-based and three-track method can achieve a similar quality per bitrate. 
Our method is limited better or worse than the other solution in specific scene or video content. 
Besides, it is worthy to note that PSNR value can be improved by about 0.5-1dB because of 
the SVC structure in Fig. \ref{fig6_2}.

On the other hand, the transport bitrate is also one of the factors we should consider. 
According to the previous discussion, our scheme can reduce the latency to one-frame time. 
If the existing method that wants to reduce the latency to one frame as well, it needs to transport 
the full picture or at least provide enough margin depending on the specified scene like \cite{akcay2021head}. 
Here we do not discuss the transmission strategy for specific video scenes, because our approach 
requires the only about 1/6 of enhanced layers and simplify the adjustment of the strategy accordingly, 
which means considerable savings in bitstream.

\subsection{Encoding Time}
We consider the encoding time of each method in different encoding bitrates. 
The result is recorded from the time command in Linux system and the result equals to 
the sum of user time and system time.

The performance between SVC-based and existing three-track method is shown in Fig. \ref{enc_time}. 
According to the results, the actual encoding time of new SVC-based coding schema is significantly less 
(>30\%) than three-track method in the same bitrate, which implies using SVC can reduce a considerable 
encoding cost in multi-rate (or multi-resolution) use cases.

\section{Conclusions}
The viewport-dependent streaming becomes necessary for 360-degree videos (and VR content). 
But the temporal dependency used in modern video codecs makes viewport switch latency correlated to 
the GOP size - difficult to achieve instant viewport change which is critical for user experiences. 
In this paper, an SVC-based and tiled-based video coding structure is proposed to resolve the viewport 
switch challenge and it can reduce interactive latency to one frame time - the best possible time. 
Secondly, by using tiles in SVC layers, it circumvents the issue of packing the non-adjacent regions, 
which is a common geometry challenge in non-planar contents, such as 360-degree videos or VR. 
Furthermore, compared with existing industry approaches, this new coding schema can save encoding 
cost significantly, while achieving a similar PSNR result in same bitrate.

On a side note, the encoding performance of AV1 spatial SVC may be an interesting topic for further study. 
In our experiments, 2x spatial scale factor is used between the base layer and enhanced layer, 
because SVC usually needs a smaller scale factor for better quality \cite{wien2007performance}. Depending on video contents, 
AV1 SVC encoding may present different quality metrics compared with conventional single layer encoding, 
but the gap (PSNR) is acceptable in our case - even in the worst case, PSNR gap is less than 1dB 
(typically ~0.5dB).

In the end, the new coding schema can be implemented with AV1 SVC \cite{de2018av1}, without changes in the existing 
codec standards - i.e., no changes in AV1 decoder. Similar approach can be also  implemented in MPEG-I VVC, 
with  subpicture and reference picture resampling (RPR)\cite{bross2021overview}. In addition, the 360-degree video delivery 
architecture specified in MPEG-I OMAF \cite{choi2017information} - such as applications, encoding codecs, and packing formats, 
can be greatly simplified by using this SVC-based coding schema. This work paves the way for the most 
efficient viewport-dependent streaming of 360-degree videos, in terms of interactive latency, 
coding performance, and architecture.

\begin{figure*}[bp]
    \centering
    \subfigure[Toyko]{
        \includegraphics[width=0.32\textwidth]{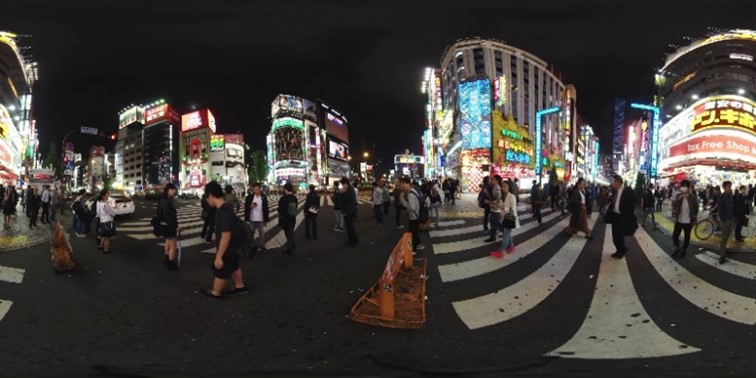}}
    \subfigure[Office]{
        \includegraphics[width=0.32\textwidth]{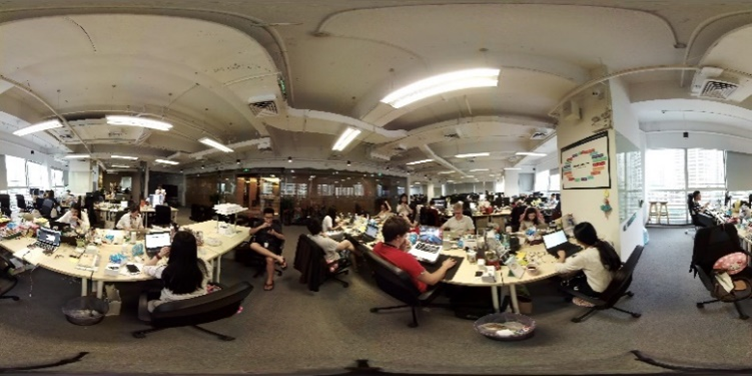}}
    \subfigure[Beach]{
        \includegraphics[width=0.32\textwidth]{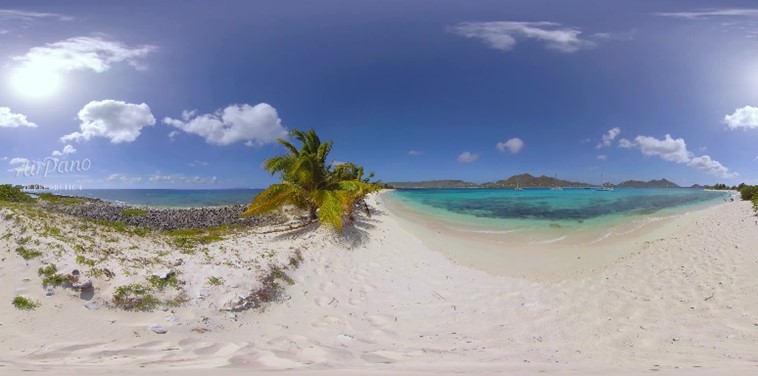}}
    \subfigure[Lion]{
        \includegraphics[width=0.32\textwidth]{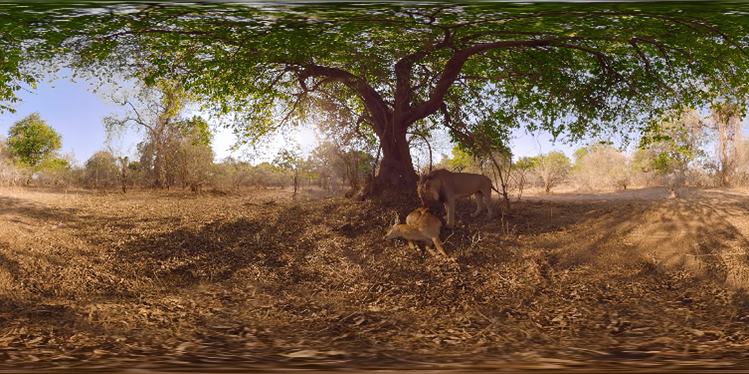}}
    \subfigure[Ski]{
        \includegraphics[width=0.32\textwidth]{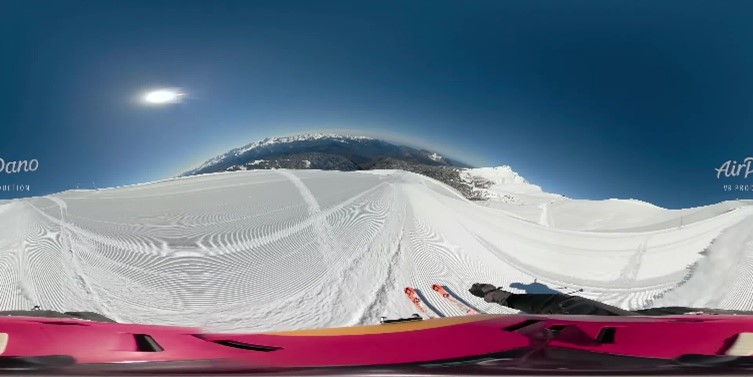}}
    \subfigure[Street]{
        \includegraphics[width=0.32\textwidth]{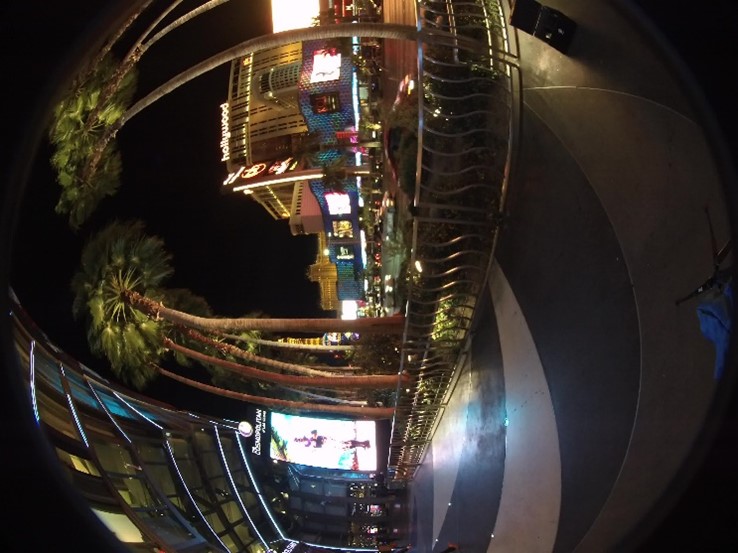}}
    \caption{PSNR under different encoding configurations}
    \label{clips}
\end{figure*}

\bibliographystyle{unsrt}
\bibliography{export}

\begin{thebibliography}{10}

\bibitem{ichigaya2016required}
Atsuro Ichigaya and Yukihiro Nishida.
\newblock Required bit rates analysis for a new broadcasting service using
  {{HEVC}}/{{H}}. 265.
\newblock {\em IEEE Transactions on Broadcasting}, 62(2):417--425, 2016.

\bibitem{hannuksela2019overview}
Miska~M Hannuksela, Ye-Kui Wang, and Ari Hourunranta.
\newblock An overview of the {{OMAF}} standard for 360 video.
\newblock In {\em 2019 {{Data}} Compression Conference ({{DCC}})}, pages
  418--427. {IEEE}, 2019.

\bibitem{choi2017information}
B~Choi, {\relax YK}~Wang, {\relax MM}~Hannuksela, Y~Lim, and A~Murtaza.
\newblock Information technology\textendash coded representation of immersive
  media ({{MPEG-I}})\textendash part 2: {{Omnidirectional}} media format.
\newblock {\em ISO/IEC}, pages 23090--2, 2017.

\bibitem{o.v.cloudImmersiveVideoSample}
O.~V. Cloud.
\newblock Immersive {{Video Sample}}.

\bibitem{tiledmediaHowClearVRDrives}
Tiledmedia.
\newblock How {{ClearVR Drives}} and {{Leverages Standards}}.

\bibitem{tiledmediaTiledmediaTiledStreaming}
Tiledmedia.
\newblock Tiledmedia {{Tiled Streaming Enables Streaming 8K 360VR}}.

\bibitem{podborski2019html5}
Dimitri Podborski, Jangwoo Son, Gurdeep~Singh Bhullar, Robert Skupin, Yago
  Sanchez, Cornelius Hellge, and Thomas Schierl.
\newblock {{HTML5 MSE}} playback of {{MPEG}} 360 {{VR}} tiled streaming:
  {{JavaScript}} implementation of {{MPEG-OMAF}} viewport-dependent video
  profile with {{HEVC}} tiles.
\newblock In {\em Proceedings of the 10th {{ACM}} Multimedia Systems
  Conference}, pages 324--327, 2019.

\bibitem{you2020omaf4cloud}
Yu~You, Ari Hourunranta, and Emre~B Aksu.
\newblock {{OMAF4Cloud}}: {{Standards-enabled}} 360\textdegree{} video creation
  as a service.
\newblock {\em SMPTE Motion Imaging Journal}, 129(9):18--23, 2020.

\bibitem{zhang2022realvr}
Qi~Zhang, Jianchao Wei, Shanshe Wang, Siwei Ma, and Wen Gao.
\newblock {{RealVR}}: {{Efficient}}, economical, and
  quality-of-experience-driven {{VR}} video system based on {{MPEG OMAF}}.
\newblock {\em IEEE Transactions on Multimedia}, 2022.

\bibitem{albert2017latency}
Rachel Albert, Anjul Patney, David Luebke, and Joohwan Kim.
\newblock Latency requirements for foveated rendering in virtual reality.
\newblock {\em ACM Transactions on Applied Perception (TAP)}, 14(4):1--13,
  2017.

\bibitem{sullivan2012overview}
Gary~J Sullivan, Jens-Rainer Ohm, Woo-Jin Han, and Thomas Wiegand.
\newblock Overview of the high efficiency video coding ({{HEVC}}) standard.
\newblock {\em IEEE Transactions on circuits and systems for video technology},
  22(12):1649--1668, 2012.

\bibitem{de2018av1}
Peter De~Rivaz and Jack Haughton.
\newblock Av1 bitstream \& decoding process specification.
\newblock {\em The Alliance for Open Media}, 681, 2018.

\bibitem{insta360Insta360Pro8K}
Insta360.
\newblock Insta360 {{Pro 8K Sample}}.

\bibitem{airpianoCaribbeanParadise360}
Airpiano.
\newblock Caribbean {{Paradise}} 360 video in {{8K}} \textendash{} {{YouTube}}.

\bibitem{geographicLions360}
N.~Geographic.
\newblock Lions 360\textdegree.

\bibitem{airpianoRosaKhutorSki}
Airpiano.
\newblock Rosa {{Khutor Ski Reso}}.

\bibitem{akcay2021head}
Mehmet~N Akcay, Burak Kara, Saba Ahsan, Ali~C Begen, Igor Curcio, and Emre
  Aksu.
\newblock Head-motion-aware viewport margins for improving user experience in
  immersive video.
\newblock In {\em {{ACM}} Multimedia Asia}, pages 1--5. 2021.

\bibitem{wien2007performance}
Mathias Wien, Heiko Schwarz, and Tobias Oelbaum.
\newblock Performance analysis of {{SVC}}.
\newblock {\em IEEE Transactions on Circuits and Systems for Video Technology},
  17(9):1194--1203, 2007.

\bibitem{bross2021overview}
Benjamin Bross, Ye-Kui Wang, Yan Ye, Shan Liu, Jianle Chen, Gary~J Sullivan,
  and Jens-Rainer Ohm.
\newblock Overview of the versatile video coding ({{VVC}}) standard and its
  applications.
\newblock {\em IEEE Transactions on Circuits and Systems for Video Technology},
  31(10):3736--3764, 2021.

\end{thebibliography}

\end{document}